\documentclass[twocolumn,showpacs,preprintnumbers,amsmath,amssymb]{revtex4}

\usepackage{graphicx}% Include figure files
\usepackage{dcolumn}% Align table columns on decimal point
\usepackage{bm}% bold math

\newcommand{\ket}[1]{\ensuremath{| #1 \rangle}}
\newcommand{\bra}[1]{\ensuremath{\langle #1 |}}

\begin{document}

\title{Analysis of Photoassociation Spectra for Giant Helium Dimers.}

\author{J. L\'eonard}
\email{leonard@lkb.ens.fr}
\author{A. P. Mosk}
 \altaffiliation[Permanent address: ]{Dept. of science \& technology, and MESA+ institute, University of Twente,
Netherlands.}
\author{M. Walhout}
 \altaffiliation[Permanent address: ]{Calvin College, Grand Rapids, MI, USA. }
\author{P. van der Straten}
 \altaffiliation[Permanent address: ]{Utrecht University, Netherlands. }

\author{M. Leduc}
\author{C. Cohen-Tannoudji}

\affiliation{Ecole Normale Sup\'erieure and Coll\`ege de France\\
Laboratoire Kastler Brossel, 24 rue Lhomond, 75231 Paris Cedex 05,
France}

\date{\today}% It is always \today, today,

\begin{abstract}
We perform a theoretical analysis to interpret the spectra of purely
long-range helium dimers produced by photoassociation (PA) in an
ultra-cold gas of metastable helium atoms. The experimental spectrum
obtained with the PA laser tuned closed to the $2^3S_1\leftrightarrow
2^3P_0$ atomic line has been reported in a previous Letter. Here, we
first focus on the corrections to be applied to the measured resonance
frequencies in order to infer the molecular binding energies. We then
present a calculation of the vibrational spectra for the purely
long-range molecular states, using adiabatic potentials obtained from
perturbation theory. With retardation effects taken into account, the
agreement between experimental and theoretical determinations of the
spectrum for the $0_u^+$ purely long-range potential well is very good.
The results yield a determination of the lifetime of the $2^3P$ atomic
state.

\end{abstract}

\maketitle

\section{Introduction}
\label{Section: Introduction}

Photoassociation (PA) spectroscopy is a powerful technique for
acquiring information about the collisional properties of laser-cooled
atoms. It has revealed a rich array of high-resolution spectroscopic
data for alkali diatomic molecules \cite{Revues} and provided a means
of testing calculations of molecular dynamics.  It has also led to good
estimates of the s-wave scattering length \cite{Abraham,Gardner} that
determines the behavior of ultra-cold dilute gases near quantum
degeneracy.

The case of $^4$He atoms in the metastable $2^3S_1$ state (He$^*$) is
distinctive in that each atom carries a large internal energy of 20 eV.
Photoassociation experiments with He$^*$ were first demonstrated by
Herschbach {\it et al.} with atoms trapped in a magneto-optical trap
(MOT) \cite{Herschbach}. However, the quantitative study of pair
interactions has still to be completed. In particular, although Bose
Einstein Condensation (BEC) has been achieved in He$^*$
\cite{Pereira101,Robert}, the scattering length remains uncertain. What
is more, the accurate investigation of collisional properties
\cite{Sirjean} and of the dynamical behavior \cite{Leduc} of the
ultra-cold He$^*$ gas suffers from the uncertainty in the scattering
length. In order to extract quantitative information from PA
spectroscopy we have performed a new PA experiment starting from a
magnetically trapped and evaporatively cooled metastable helium gas. We
have thereby achieved greater state selectivity, higher density, and
lower temperature than were obtained previously \cite{Herschbach}.

\begin{figure}[t]
\begin{center}
\includegraphics[height=7cm]{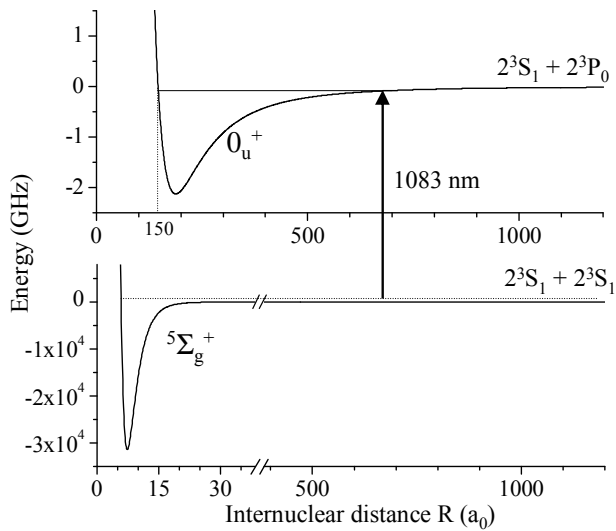}
\end{center}
\caption{\footnotesize a) Illustration of the principle of a
photoassociation (PA) experiment. A free pair of metastable atoms is
resonantly excited into a purely long-range $0_u^+$ molecular bound
state. The potential curve for the $^5\Sigma_g^+$ state is the one
given by \cite{StarckMeyer}, the $0_u^+$ is the one obtained by the
calculation described in the text. Note the change in energy and length
scales between the $^5\Sigma_g^+$ and the purely long-range $0_u^+$
potential wells.}\label{Fig: Principe_PA}
\end{figure}

As a preliminary step toward the characterization of pair interactions,
we have reported \cite{Leonard} the observation of purely long-range
helium dimers produced by photoassociation of metastable helium atoms,
with the PA laser tuned close to the $2^3S_1 \leftrightarrow 2^3P_0$
atomic line (see Figure \ref{Fig: Principe_PA}). The novelty of these
dimers is that they are produced from two highly excited atoms and
therefore carry a huge internal energy of 40 eV. However, whereas one
might expect the molecules to decay through autoionization, the primary
decay mechanism is radiative.  This fact allowed us to develop an
original, ``calorimetric'' detection method based on the strong heating
of the atomic cloud at resonant PA frequencies. Our preliminary model
for the heating accounts for the conversion of a decaying molecule's
vibrational kinetic energy into additional thermal energy within the
cloud. Autoionization appears to have a negligible effect, probably
because the inner turning points for these giant dimers are so far
apart (around 150 bohr radii).  Ionization is unlikely at such
distances, so it is not surprising that these molecular states have not
been observed with the ion detectors used in MOT experiments
\cite{Herschbach}.

The present paper is meant to provide a theoretical complement to
reference \cite{Leonard}, which focused primarily on experimental
methods and results. Because $^4$He has no hyperfine structure, the
theoretical approach is relatively simple as compared with alkali
systems. Thus, giant helium dimers present an interesting case study,
and we have attempted to emphasize important physical concepts in
somewhat of a tutorial approach. In particular, a perturbative
description of the electronic potentials is given, which provides a
physical understanding of the formation of these molecules. Then, with
a single-channel adiabatic calculation of the effective molecular
potentials we find purely long-range spectra that are in excellent
agreement with those computed in \cite{Venturi} by more sophisticated
techniques.

In Section \ref{section: shifts}, after a brief review of the
experiment, we relate the molecular binding energy to the measured
resonance frequency by subtracting shifts due to the magnetic trapping
potential and the non-zero temperature of the atomic cloud. In
particular, the free-bound character of the transitions leads to
temperature-induced shifts which do not exist in the case of
bound-bound transitions. Section \ref{section: RoVib structure}
describes the calculation of the long-range part of the $2^3S$ - $2^3P$
molecular interaction potentials, as well as the theoretical values for
the binding energies of the giant dimers. Our perturbative approach
shows how purely long-range potential wells arise from the competition
between the dipole-dipole interaction and the atomic fine structure.
Finally, we compare both the experimental and theoretical
determinations of the binding energies. With its high accuracy, the
experiment provides a clear illustration of retardation effects in the
electromagnetic interaction and of tiny corrections due to the
vibration-induced coupling between electronic and nuclear degrees of
freedom. Moreover, it yields a measurement of the radiative decay rate
$\Gamma$ of the atomic excited state $2^3P$ with an accuracy of
$0.2\%$.

\section{Deriving the binding energies from PA measurements}
\label{section: shifts}

\subsection{Acquisition of PA spectra}

\begin{figure}[t]
\begin{center}
\includegraphics[height=6.5cm]{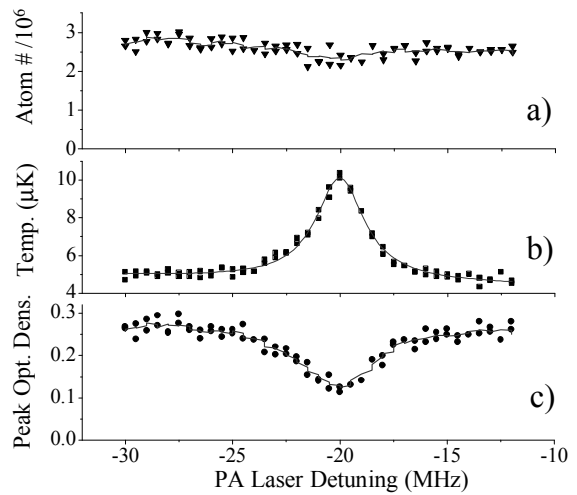}
\end{center}
\caption{\footnotesize Detection of the resonant formation of giant
dimers in the $v=4$ vibrational state of the $0_u^+$ potential well.
After the PA laser pulse and further thermalization, the remaining
atoms are detected optically: a) atom number, b) temperature in $\mu$K
and c) peak optical density versus the PA laser detuning from the
atomic $D_0$ line. Each point represents a new evaporated cloud after
PA pulse illumination, thermalization and ballistic expansion. The
curves in graphs a) and c) indicate the averaging of data over 5
adjacent points. The curve in graph b) is a Lorentzian fit to the data
with a width of 2.8 MHz. Strong heating of the atomic cloud is
observed when the PA laser is resonant with a molecular transition.}
\label{Fig: Data_v4}
\end{figure}

We perform PA experiments with a cold metastable helium gas confined in
a magnetic trap. The atomic cloud is cooled by RF-induced evaporation
to a temperature in the $\mu$K range, just above the BEC transition
\cite{Pereira2}.  The cloud is illuminated for a short period (0.1 to
10 ms) by a low-intensity PA laser beam and then allowed to thermalize
for a few hundred ms.  It is subsequently released and then detected
optically after a few-ms expansion time.  Giant helium dimers are
produced when a free (unbound) pair of cold atoms absorbs a PA photon
and is excited into a bound state of the purely long-range potential.
This free-bound transition occurs when the PA laser is tuned red of the
$2^3S_1 \leftrightarrow 2^3P_0$ ($D_0$) atomic line (see Figure
\ref{Fig: Principe_PA}).  Several resonance lines appear in the
recorded temperature data, indicating that the formation of transient
molecules results in the deposition of energy in the surrounding atomic
cloud. Figure \ref{Fig: Data_v4} illustrates the typical data obtained
when we tune the PA laser in the vicinity of a bound state in the
$0_u^+$ potential well.  Although few atoms are lost (Figure \ref{Fig:
Data_v4}-a), a strong increase in temperature (Figure \ref{Fig:
Data_v4}-b) and consequently a strong decrease in peak optical density
(Figure \ref{Fig: Data_v4}-c) are monitored. Since the cloud is very
cold (typically $5 \mu$K), the excitation of relatively few molecules
is enough to cause significant heating. Thus, the atomic cloud serves
as a sensitive calorimeter capable of detecting the position of the
molecular lines with an accuracy of 0.5 MHz. The quantitative study of
the heating mechanism is in progress and will be published in a
separate paper.

\subsection{Discussion of the various line shift mechanisms}

Acquiring experimental spectra consists in measuring the PA laser
detunings at which molecular lines are resonantly excited in the
magnetically trapped atomic cloud.  For an accurate
interpretation of the data, we need to take into account the correct lineshape
function, which may include shifts and/or asymmetric broadening due
to various mechanisms.  We do so on the basis of the following calculation
of the molecular binding energy, which emerges straightforwardly from the
conservation of energy and momentum.

\subsubsection{Conservation of energy for a free-bound transition}

The energy $E_{i}$ of a pair of trapped atoms in the initial unbound state
can be written:
\begin{equation}\label{eq: Initial Energy}
E_{i}(\vec{r_1},\vec{r_2},\vec{P},\vec{p}_{rel})=\frac{\vec{P}^{\;2}}{4m}+\frac{
\vec{p}_{rel}^{\;2}}{m} - \vec{\mu}\cdot\left( {\vec{B}(\vec{r_1}) +
\vec{B}(\vec{r_2})} \right),
\end{equation} where $m$ is the mass of the He atom, $\vec{P}=\vec{p_1}+\vec{p_2}$
is the momentum of the pair's center of mass, $ \vec{p}_{rel}
=(\vec{p_1}-\vec{p_2})/2$ is the relative momentum,
${\vec{B}(\vec{r_1})}$ and ${\vec{B}(\vec{r_2})}$ are the magnetic
field at the location of each atom, and ${\vec{\mu}}$ is the magnetic
dipole moment of an atom in the $2^3S_1$ state (the Land\'e factor
being 2, we define $\mu =- 2 \mu_B$, with the Bohr magneton $\mu_B<0$).
In expression (\ref{eq: Initial Energy}), we neglect any interaction
energy between the two atoms. This will be justified below.

After the pair of atoms absorbs a photon with momentum $\hbar\vec{k}$
and frequency $\nu_L$, the binding energy $hb<0$ of the resulting
molecule can be inferred from the conservation law for energy and
momentum:
\begin{eqnarray}\label{eq:conservation}
E_i(\vec{r_1},\vec{r_2},\vec{P},\vec{p}_{rel})+ h \nu_L & =&
\frac{\vec{P}_M^{\;2}}{4m} + h(\nu_0+b)\ \\
\mbox{with }\ \vec{P} +\hbar \vec{k}& =& \vec{P}_M\ ,\nonumber
\end{eqnarray}
where $\vec{P}_M$ is the final momentum of the molecule and $h\nu_0$ is
the energy of the $D_0$ line (for an isolated atom in a zero magnetic
field). The difference between the molecular binding energy and the PA
laser detuning $\delta=\nu_L-\nu_0<0$ is thus given by:
\begin{eqnarray}  \label{eq:binging energy}
h(b-\delta) &=&  - \hbar \vec{k}\cdot
\frac{\vec{P}}{2m} - \frac{\hbar^2k^2} {4m} \nonumber\\
&&- \vec{\mu}\cdot\left( {\vec{B}(\vec{r_1}) + \vec{B}(\vec{r_2})}
\right) + \frac{ \vec{p}_{rel}^{\;2}}{m} ,
\end{eqnarray}
Any dependence of the molecular level energy on the magnetic field
(Zeeman effect) or on the density (mean field interaction of the
molecule with the surrounding atomic and/or molecular cloud) is {\it a
priori} included in $b$, which may therefore also depend on the
position of the molecule.

Note that the relative kinetic energy term $\vec{p}_{rel}^{\;2}/m$ in
Equation (\ref{eq:binging energy}) would not appear in the case of a
bound-bound transition, since it would be implicitly included in the
initial binding energy. As it is always positive, it contributes an
{\it asymmetric} lineshape, and consequently a mean shift
\cite{Napolitano}. Also, the harmonic magnetic trapping potential
contains quadratic terms which contribute to the inhomogeneous,
asymmetric broadening and shift of the lines. However, the temperature
is low enough that the asymmetric broadening terms remain much smaller
than the natural lorentzian width. Thus, the only effect is a shift of
the peak position of the lines, which can be calculated by averaging
Equation (\ref{eq:binging energy}) over the distribution function for
the initial pair of free atoms.

\subsubsection{Initial distribution function of the free pair}

The distribution function for the pairs {\it that undergo the PA
transition} is the thermal distribution for a pair of trapped atoms
{\it multiplied} by the transition probability.  According to the
Franck-Condon overlap principle, the latter is proportional to the
square of the overlap between the initial and final radial wave
functions. Since the excited state is a bound state, the overlap is
peaked at the Condon radius $R_C$ close to the classical outer turning
point. According to Table \ref{tab: results} in Section \ref{section:
RoVib structure}, the transition occurs mainly for an internuclear
distance $R_C=||\vec{r_1}-\vec{r_2}|| \lesssim 50 $ nm, which is much
smaller than the size of the atomic cloud ($\sim 100\ \mu$m at $T\sim
10\ \mu$K). This allows us to use the approximation
$\vec{r_1}\simeq\vec{r_2}\simeq\vec{r}$ in Equations (\ref{eq: Initial
Energy}) and (\ref{eq:binging energy}), where $\vec{r}$ is the center
of mass of the pair.  Furthermore, because the temperature is so low,
the collision between two atoms occurs in the s-wave scattering regime,
for which the relative angular momentum $\vec{p}_{rel}=\hbar \vec{q}$
has no component orthogonal to the internuclear axis. Thus, the
vectorial character of $\vec{p}_{rel}$ can be ignored, since there is
only one degree of freedom for the relative motion of the colliding
atoms. For internuclear distances $R$ close to $R_C$, the radial part
$u(R)$ of the ground state wave function can be approximated as
$u(R)\propto \sin(q(R-a))\propto q$ since $qR_C \ll 1$ (with $a$
representing the s-wave scattering length; see {\it e.g.}
\cite{Revues}). Finally, the distribution function for a pair of
trapped atoms in the s-wave scattering regime absorbing a PA photon is
found to be proportional to:
\begin{eqnarray}\label{eq:distrib}
q^2 \delta(\vec{r_1}-\vec{r_2})\times
\exp(-E_{i}(\vec{r_1},\vec{r_2},\vec{P},q)/k_BT)\ .
\end{eqnarray}

\subsubsection{Mean frequency shifts}

i)Average over the center-of-mass momentum.\\

The first term in the right-hand side of Equation (\ref{eq:binging
energy}) is responsible for the Doppler profile. It produces no average
shift, since there is \textit{a priori} no correlation between the
momenta of the two atoms and of the photon: $\langle
\vec{k}.\vec{P}\rangle=0$. However, it is responsible for a symmetric
broadening of the lines, which scales like $\sqrt{T}$ ($T$, the
temperature of the cold gas). In the microK range of temperature, this
Doppler broadening turns out to be small compared with the natural
lifetime broadening of the molecular states probed.

The second term in Equation (\ref{eq:binging energy}) is the recoil
energy of the molecule after absorbing the photon. In units of $h$, its
numerical value is $\sim 21$ kHz, which is well below our experimental
accuracy. Therefore we neglect the corresponding shift.\\

ii) Average over the center of mass position.\\

Using expression (\ref{eq:distrib}), the average over the positions
$\vec{r_1}$ and $\vec{r_2}$ turns out to be an average over the
position $\vec{r}$ of the center of mass of the pair. The shift induced
by the external trapping potential is thus calculated to be:
\begin{equation}
\left\langle - \vec{\mu}\cdot\left( {\vec{B}(\vec{r_1}) +
\vec{B}(\vec{r_2})} \right) \right\rangle = 2 \mu B_0 +
\frac{3}{2}k_BT,
\end{equation}
where $2 \mu B_0$ is twice the Zeeman shift of one atom at the center
of the trap, and $3 k_BT/2$ is the average of the harmonic trapping
potential energy, according to the equipartition theorem for quadratic
energy terms.

As already noted, the binding energy $hb$ {\it a priori} also depends
on the center of mass position, and should therefore be averaged as
well. However, we neglect this position dependence, since the effect of
both the inhomogeneous magnetic field (molecular Zeeman effect) and
density (atom-molecule interaction) turn out to be small compared with
our experimental accuracy, as discussed below.
\\

iii) Average over the relative momentum.\\

Making use of expression (\ref{eq:distrib}), we find the average of the
relative kinetic energy term:
\begin{eqnarray}
\left\langle \frac{\hbar^2 q^2}{m} \right\rangle =
\frac{\int\frac{\hbar^2 q^2}{m}\  q^2 \exp (-\frac{\hbar^2
q^2}{mk_BT})\;dq}{\int q^2 \exp (-\frac{\hbar^2
q^2}{mk_BT})\;dq}=\frac{3}{2}k_BT,
\end{eqnarray}
where the denominator normalizes the distribution function. Let us
mention that while there is only one degree of freedom for the relative
momentum (in the s-wave scattering regime), our inclusion of the pair
distribution function leads us coincidentally to the same $3k_BT/2$
that one finds when treating three classical degrees of freedom.
\\

iv) Other shift mechanisms.\\

The mean-field interaction due to the surrounding medium on both the
initial and final states of the transition can cause density-dependent
shifts of the lines. As far as the initial pair of free atoms is
concerned, the mean field interaction energy is $4\pi\hbar^2\times
na/m$, where the atomic density $n<10^{14}$ cm$^{-3}$, the s-wave
scattering length $a<20$ nm \cite{Pereira101,Robert}, and $m\sim
6.68\times 10^{-27}$ kg. In units of $h$, the upper bound for this
mean-field interaction is less than $\sim 60$ kHz, which is below our
experimental accuracy and therefore negligible.  The mean field energy
shift of the final molecular state, which would appear as a
density-dependent term in the experimental binding energy, has not been
detected experimentally.

Finally, light-induced line shifts are completely negligible, since the
spectra were measured with PA laser intensities well below the atomic
saturation intensity.
\\

v) Summary.\\

In our experiment, each molecular line produces a resonant increase in
temperature as a function of PA detuning $\delta$.  Each resonance line
is fit by a Lorentzian. The fit's center frequency $\delta_v$ is taken
to be the resonant frequency for excitation to vibrational level $v$.
Accounting for the corrections described above, we infer the molecular
binding energy $hb_v$ of this vibrational level to be:
\begin{equation}
h b_v \simeq h \delta_v + 2\mu B_0 + 3k_BT. \label{eq:e3}
\end{equation}

\subsection{Experimental checks for the lineshifts}

We have measured $\delta_v$, $B_0$ and $T$ for the lines $v=0$ through
$v=4$ in the $0_u^+$ potential well, for $B_0=0.1$ to $\sim10 $ Gauss
and for $T= 1.5$ to 30 $\mu$K. The temperature of the gas was varied by
changing the final RF frequency of the evaporation ramp above the
critical temperature. Consequently, the atomic density was also varied
from $n\sim0.5\times 10^{13}$ to $\sim 8 \times 10^{13}$ at$/$cm$^{3}$.

\begin{figure}[b]
\begin{center}
\includegraphics[height=6cm]{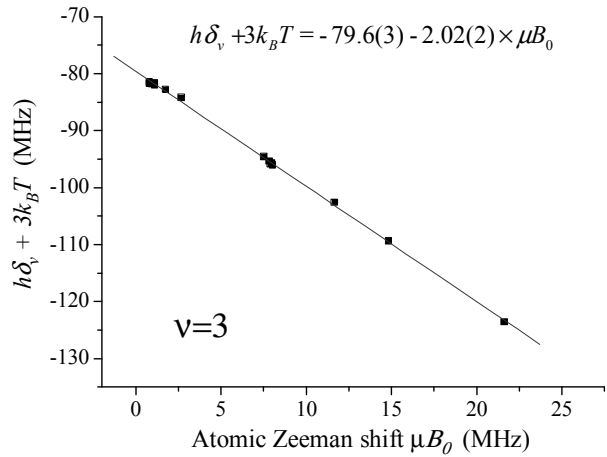}
\end{center}
\caption{\footnotesize Experimental determination of the binding
energy in the $0_u^+$ potential well for the vibrational level $v=3$:
illustration of the dependence of the measured detuning $\delta_v$ on
the magnetic field $B_0$, after correction from the temperature-induced
shift (see Equation (\ref{eq:e3})). The slope of the linear fit is
compatible with the expected dependence in $B_0$ (see in the text).}
\label{Fig: B0_dependence}
\end{figure}

In Equation \ref{eq:e3} the most important correction is due to $B_0$.
Figure \ref{Fig: B0_dependence} shows the dependence on $B_0$ of the
measured detuning $\delta_v$ of the $v=3$ line, after it is corrected
for the temperature-induced effect ($3k_BT$). If the magnetic field is
measured in units of $\mu B_0$, a linear fit to the data gives a slope
of $-2.02\pm0.02$. Given Equation (\ref{eq:e3}), the contribution of
the initial pair of free cold atoms (the ``ground" state), should be
exactly $-2 \mu B_0$. A deviation from this value could be attributed
to the contribution of the mean Zeeman effect of the molecular bound
(``excited") state. As the $0_u^+$ electronic state is non degenerate,
the molecule cannot have any magnetic dipole moment except one induced
by the molecular rotation, which is expected to be of the order of the
nuclear magneton, or about three orders of magnitude smaller than
$\mu_B$. Given the experimental accuracy and the range of magnetic
field explored, the correspondingly small Zeeman effect would be
difficult to measure. But our data permit us to set an upper bound of
$0.02\; \mu = 0.04\;|\mu_B|$ on the molecular magnetic dipole moment.
This result justifies neglecting the molecular Zeeman effect in the
calculation of the mean line shifts.

\begin{figure}[b]
\begin{center}
\includegraphics[height=6cm]{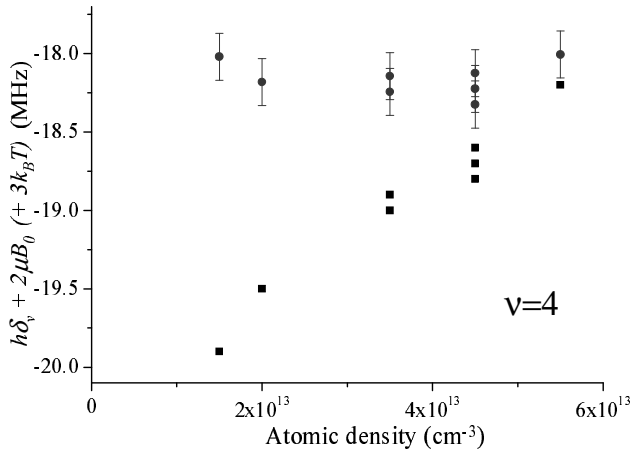}
\end{center}
\caption{\footnotesize Experimental determination of the binding
energies in the $0_u^+$ potential well: illustration, in the case of
the vibrational level $v=4$, of the dependence of the measured detuning
$\delta_v$ on the temperature and on the density, after correction from
the magnetically-induced shift (see Equation (\ref{eq:e3})). Data are
displayed before (squares) and after (circles) applying the
temperature-dependent correction. Error bars include uncertainty in the
measurements of $\delta$, $B_0$, and $T$.} \label{Fig: T_n_dependence}
\end{figure}

Figure \ref{Fig: T_n_dependence} displays the measured position of the
$v=4$ line, corrected for the magnetically-induced shift ($2\mu B_0$),
as function of the atomic cloud density. Data with (circles) and
without (squares) the additional temperature-dependent correction are
shown. The uncorrected data has been displayed in order to illustrate
the importance of the temperature effect (up to 2 MHz at $\sim 30\
\mu$K) as compared to the experimental accuray (0.5 MHz). For this set
of data, the density was increased simply by further evaporative
cooling of the gas. Thus, higher density is associated with lower
temperature, and the temperature-induced shift indicated by the squares
nearly vanishes for large density. It should be noted here that the
size of the molecules ($917\  a_0 \sim 50$ nm, see Table \ref{tab:
results}) is not vanishingly small compared with mean inter-atomic
distance in the cloud ($\sim 260$ nm at $6 \times 10^{13}$
at/cm$^{3}$). Under these conditions, one might expect to find a
density- dependent shift due to the mean field interaction between the
molecule and the surrounding atomic medium. However, no such shift is
evident in our data after we apply the corrections for temperature and
magnetic field. The error bars include experimental uncertainty in
$\delta$, $B_0$ and $T$. Additional scatter of about $ 0.3$ MHz can be
attributed to the uncertainty in the PA laser frequency lock. We have
studied the stability of the experiment and the possible sources of
systematic error in all achievable parameter ranges (accumulating many
more data than are shown in Figure \ref{Fig: T_n_dependence}). We
conclude that the binding energy for $v=4$ is $-18.2\pm 0.5$ MHz, in
units of $h$.

Finally, from Figure \ref{Fig: T_n_dependence} and from the 0.5 MHz
uncertainty, we can infer that the density-induced energy shift of the
molecules must be smaller than $\sim 100$ kHz per $10^{13}$ cm$^{-3}$
of density. Actually, the atomic Bose gas surrounding the molecule is
near resonance and therefore has a permittivity that differs from the
vacuum value. For an ideally homogeneous medium, the permittivity would
enter in the resonant dipole potential \cite{Dip-Dip_in_medium},
leading to a density-dependent term in the binding energy which would
be at least a factor two above our upper bound. Since we do not detect
this effect, we conclude our gas can not be considered as an
homogeneous medium on the size scale of a molecule. This point may
deliver important information about the three-particle correlation
function in the atomic gas and would require further study, but it has
not been investigated so far.

Similar data were registered for the other vibrational lines that we
were able to measure. The experimental results for the binding energies
are reported in Table \ref{tab: comparison}, Section \ref{section:
RoVib structure}.

\section{Ro-vibrational structure of the giant dimers}
\label{section: RoVib structure}

In order to interpret the measurements described above, we now develop
the calculation of the long-range interaction of one atom in the
$2^3S_1$ state, and another one in the $2^3P_{J=0,1,2}$ state. It
happens that some of the resulting potential energy curves have minima
at very large internuclear distance and support purely long-range bound
states.  In particular, the calculated spectrum of five vibrational
states in the $0_u^+$ potential will be shown to be in excellent
agreement with our measurements.

\subsection{Electronic potential curves for the $2^3S+2^3P$ system with fixed nuclei}
\label{section: adiabatic potentials}

\subsubsection{Hamiltonian}

The general task for calculating molecular potentials in $^4$He
consists in solving the following Schr\"odinger equation
\cite{Lefebvre-Brion}:
\begin{eqnarray}
\label{Hamiltonian_General}
&&\hat{H}\ket{\psi_{\alpha}}=(\;\hat{T}_n+\hat{T}_e+\hat{V}+\hat{H}_{rel}\;)\ket{\psi_{\alpha}}=E_{\alpha}\ket{\psi_{\alpha}}\\
&&\mbox{where  }\hat{T}_n=\displaystyle \sum_{k=1}^2
\frac{\hat{\textbf{p}}_\textbf{k}^2}{2M}\mbox{ ,    }\hat{T}_e
=\displaystyle \sum_{i=1}^4
\frac{\hat{\textbf{p}}_\textbf{i}^2}{2m}\mbox{ ,} \nonumber\\
&&\mbox{and  }\hat{V}=\hat{V}(\hat{\textbf{r}}_k,\hat{\textbf{r}}_i)
\mbox{ , } \hat{H}_{rel} =
\hat{H}_{rel}(\hat{\textbf{r}}_i,\hat{\textbf{s}}_i).\nonumber~
\end{eqnarray}
Here, $\ket{\psi_{\alpha}}$ is a stationary solution corresponding to a
set of quantum numbers $\{\alpha\}$ to be detailed later. The
hamiltonian written above appears as the sum of four terms $\hat{T}_n$,
$\hat{T}_e$, $\hat{V}$ and $\hat{H}_{rel}$ which represent respectively
the kinetic energy of the two nuclei, the kinetic energy of the four
electrons, the non relativistic interaction between the six charged
particles, and the relativistic part of the hamiltonian. This operator
is written as function of the positions of the nuclei
$\hat{\textbf{r}}_k$, and of the electrons $\hat{\textbf{r}}_i$, and as
function of the spin coordinates $\hat{\textbf{s}}_i$ of the four
electrons. The $^4$He nuclei have no spin. To solve this very
complicated problem, we adopt a perturbative approach, in which we
consider the internuclear distance large enough that the interaction
potential $\hat{V}$ can be treated as a perturbation of the system of
two independent atoms $A$ and $B$. Thus the hamiltonian
(\ref{Hamiltonian_General}) is approximated as follows:
\begin{eqnarray}
\label{Hamiltonian_approx} \hat{H}=\;\hat{T}_n + \hat{H}_{0}(A) +
\hat{H}_{0}(B) + \hat{H}_{fs}(A) + \hat{H}_{fs}(B) + \hat{U}(R)
\end{eqnarray}
where $\hat{H}_{0}$ and $\hat{H}_{fs}$ are respectively the
non-relativistic and relativistic part of the hamiltonian for one
isolated atom, and $\hat{U}(R)$ stands for the long-range electrostatic
interaction between the two atoms, whose leading term is the retarded
dipole-dipole interaction.

To describe long-range molecular interactions, we expand the molecular
state in linear combinations of (entangled) atomic states (LCAO
approximation). Moreover, according to the usual Born-Oppenheimer
approximation we first consider only the electronic degrees of freedom
while keeping the nuclei (more precisely, the atomic
centers of mass) fixed.  We then treat both the dipole-dipole
interaction and the atomic fine structure as perturbations of the
non-relativistic hamiltonian for two independent atoms.  We write the
two interactions in the basis set of states formed by the tensorial
product of isolated non relativistic atomic states: $\{\ket{\mbox{atom
} A: L_A,M_{LA};S_A,M_{SA}}\otimes \ket{\mbox{atom } B:
L_B,M_{LB};S_B,M_{SB}}\}$.  Considering one atomic orbital in the $2^3S$
state and another one in the $2^3P$ state, the space of states is of
dimension 54. As the two nuclei are identical, the hamiltonian is
unchanged under the inversion $\hat{I}_e$ of all the electrons with
respect to the center of mass \cite{Herzberg}. The operator $\hat{I}_e$
commutes with the hamiltonian (\ref{Hamiltonian_approx}) and has two
eigenvalues $\omega=\pm 1$ with eigenstates labeled {\it gerade} ($g$)
and {\it ungerade} ($u$) respectively.

\subsubsection{Retarded dipole-dipole interaction}
The dipole-dipole interaction $\hat{U}(R)$, first, only couples the
orbital angular momenta of the two independent non-relativistic atoms.
It is diagonal in the Hund's case (a) basis set labelled
$\ket{^{2S+1}\Lambda_{u/g}}$ (see e.g. \cite{Herzberg,Dashevskaya}).
These states can be written as follows in the atomic basis:
\begin{eqnarray}
\label{Base_du_cas_a}
\ket{^{2S+1}\Lambda_{u/g}}=\frac{1}{\sqrt{2}}\;(\;1+\omega\hat{I}_e\;)\ket{A:0,0;B:1,M_L}\otimes \ket{S,M_S}\nonumber~\\
=\frac{1}{\sqrt{2}}(\ket{A:0,0;B:1,M_L}-\omega(-1)^S\ket{A:1,M_L;B:0,0})\nonumber~\\
\otimes\ket{S,M_S}.\nonumber~
\end{eqnarray}
Here, $S$ is the total electronic spin of the molecule ($S=$0, 1 or 2),
$\Lambda$ is the projection onto the molecular axis of the electronic
orbital angular momentum of the molecule. In the Hund's case (a) basis,
the retarded dipole-dipole interaction is respectively given by
\cite{Dashevskaya,Meath}:
\begin{subequations}\label{eq:_Retarded_Dipole_All}
\begin{equation}\label{eq:_Retarded_Dipole_1}
-2\omega (-1)^S C_{3}/R^{3} \times\left(\cos(kR)+kR\sin(kR)\right),
\end{equation}
\begin{equation}\label{eq:_Retarded_Dipole_2}
\omega (-1)^S C_{3}/R^{3}
\times\left(\cos(kR)+kR\sin(kR)-(kR)^2\cos(kR)\right)\mbox{,}
\end{equation}
\end{subequations}
for $^{2S+1}\Sigma_{u/g}$ states (\ref{eq:_Retarded_Dipole_1}), and
$^{2S+1}\Pi_{u/g}$ states (\ref{eq:_Retarded_Dipole_2}). The
coefficient $C_3$ is related to the atomic dipole matrix element
$d=<2^3P|\hat{d_z}|2^3S>$, and thus to the radiative life time $1/
\Gamma$ of the atomic transition:
\begin{equation}\label{eq: C3}
C_3=\frac{|d|^2}{4 \pi \varepsilon_0}=\frac{3}{4}\; \hbar\Gamma \left(
\frac{\lambda}{2\pi}\right)^3,
\end{equation}
with $\varepsilon_0$ the vacuum permittivity. The fine structure
splitting is small enough that we assume the three atomic lines of
interest ($2^3S_1 \leftrightarrow 2^3P_{J=0,1,2}$) have the same
wavelength $\lambda = 1083.3$ nm within 0.1 nm.  The radiative
decay rate $\Gamma = 2\pi \times 1.6248$ MHz can be inferred from
$\lambda^2$ and from an accurate calculation of the oscillator strength
of the atomic transition \cite{Drake}. Finally, $C_3$ is found to be
$C_3=6.405$ atomic units, within a relative uncertainty of $5 \times
10^{-4}$.

\subsubsection{Fine structure coupling}
We next consider the relativistic part of the hamiltonian,
$\hat{H}_{fs}(A) + \hat{H}_{fs}(B)$, which is diagonal in the Hund's
case (c) basis (by definition of Hund's case (c), see e.g.
\cite{Herzberg}), with three eigenvalues corresponding to the three
states $2^3S_1 + 2^3P_{J=0,1,2}$. The eigenstates can only be
characterized by the projection $\Omega$ of the total electronic
angular momentum (orbital and spin) on the molecular axis
\cite{Herzberg}. In $^4$He the atomic fine structure can be modeled
using the following operator:
\begin{eqnarray}
\hat{H}_{fs}=\alpha\vec{L}.\vec{S}+\beta(\vec{L}.\vec{S})^{2},
\end{eqnarray}
where $\vec{L}$ and $\vec{S}$ are the atomic orbital and spin angular
momenta. In addition to the usual spin-orbit coupling, spin-spin
magnetic dipole interaction between the two electrons is an important
effect in helium \cite{Bethe_Salpeter}, leading to a non equidistant
splitting of the fine structure levels. In our model, the constants
$\alpha$ and $\beta$ are determined phenomenologically, in order to
reproduce the fine structure splittings which have been measured
\cite{FineStructSplit1,FineStructSplit2} very accurately:
\begin{equation}
\alpha=-\frac{\Delta_{J=2\leftrightarrow1}}{2\hbar^2}\ \ \ \mbox{ and
}\ \ \ \beta=\frac{2 \Delta_{J=1\leftrightarrow0} -
\Delta_{J=2\leftrightarrow1}} {6\hbar^4}\ \ , \nonumber \end{equation}
\begin{equation}\mbox{with }\left\{
\begin{tabular}{l}
$\Delta_{J=2\leftrightarrow1} = h \times 2.291175 $ GHz\\
$\Delta_{J=1\leftrightarrow0} = h \times 29.616950$ GHz
\end{tabular}
\right.\nonumber
\end{equation}

\begin{figure}[t]
\begin{center}
\includegraphics[height=13cm]{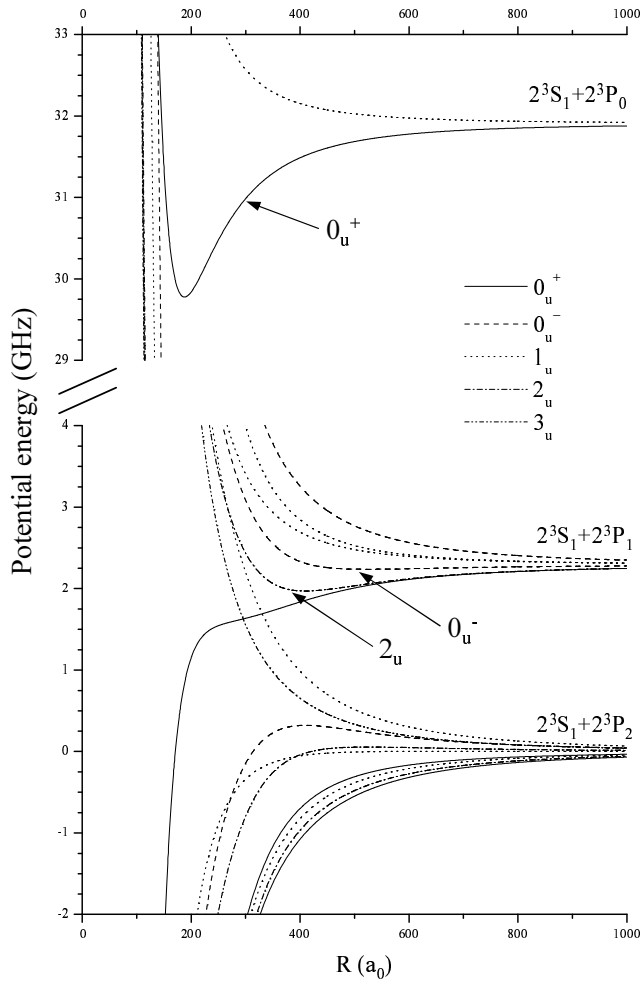}
\end{center}
\caption{\footnotesize Ungerade electronic potential curves (in GHz)
for fixed nuclei for the $2^3S+2^3P$ system versus the internuclear
distance $R$ (in atomic units; 1 $a_0 \sim 0.0529$ nm). The potential
curves result from the numerical diagonalization of the hamiltonian
(\ref{eq: non_rotating_electronic_hamiltonian}). Three arrows indicate
the three purely long-range potential wells in which bound states are
determined numerically.} \label{fig:Non_Rotating}
\end{figure}

\subsubsection{Potential curves with fixed nuclei}

According to the Movre-Pichler approach \cite{MovrePichler}, both
retarded dipole-dipole interaction and atomic fine structure coupling:
\begin{eqnarray}\label{eq: non_rotating_electronic_hamiltonian}
\hat{H}_{fs}(A) + \hat{H}_{fs}(B) +  \hat{U}(R)
\end{eqnarray}
should be considered simultaneously as a perturbation of the non
relativistic hamiltonian for two independent atoms $\hat{H}_{0}(A) +
\hat{H}_{0}(B)$. Only the projection $\Omega$ of the total electronic
angular momentum on the molecular axis is a good quantum number. States
of different $u/g$ symmetry are uncoupled and two sets of potential
curves can be determined independently for {\it gerade} and {\it
ungerade} states. Since we do photoassociation experiments in a
magnetically trapped atomic cloud, the initial quasi-molecular state is
$^5\Sigma_g^+$, and {\it gerade} states are not accessible by
single-photon excitation. Thus we focus only on {\it ungerade} states.
Figure \ref{fig:Non_Rotating} shows the results of the calculated {\it
ungerade} eigenvalues of the operator (\ref{eq:
non_rotating_electronic_hamiltonian}) as a function of $R$. Here, the
electronic states are determined with fixed nuclei. Also, the potential
curves describe only the long-range part of the molecular interactions
as a consequence of the perturbative description. For the $\Omega=0$
space, the reflection symmetry (in a plane containing the molecular
axis) leads to a relevant additional label $+/-$, which distinguishes
two states with different energies. For $\Omega\neq 0$ states, this
symmetry can be defined as well but the two resulting states have the
same energy.

\begin{figure*}[t]
\begin{center}
\includegraphics[height=6cm]{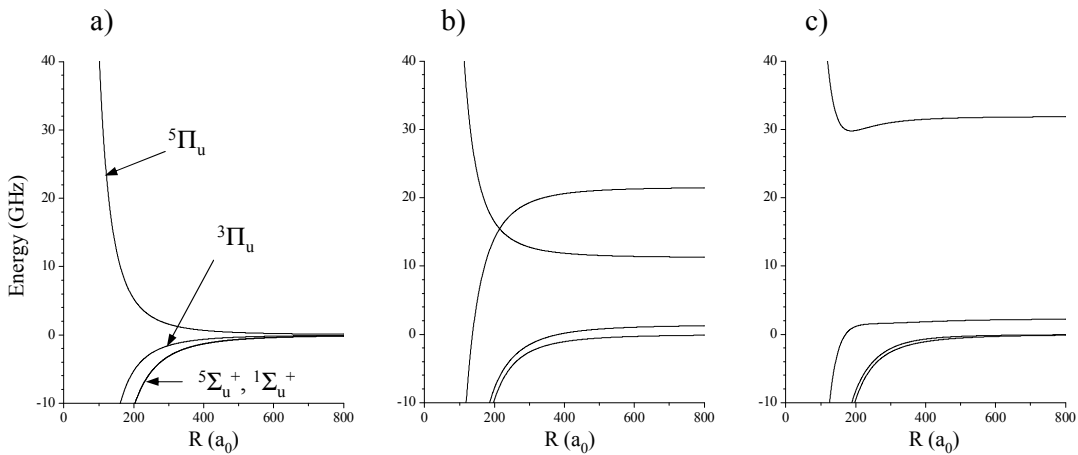}
\end{center}
\caption{\footnotesize Eigenvalues of the restriction of the
hamiltonian (\ref{eq: non_rotating_electronic_hamiltonian}) to the
$0_u^+$ subspace. Energies are given in GHz, distances are in atomic
units. a) The fine structure coupling is neglected: the eigenstates are
pure Hund's case(a) states. b) The fine structure coupling is partly
included: couplings between the repulsive state and the attractive ones
are neglected. After diagonalization, one attractive and one repulsive
states cross each other. c) Finally, including all the fine structure
coupling terms leads to an anti-crossing and a purely long-range
potential well. Note that graph b) is only for illustration and that
the neglected terms are not small.} \label{fig:
Ilustration_AntiCrossings}
\end{figure*}

\subsubsection{Physical origin of the purely long-range molecules}

The hamiltonian (\ref{eq: non_rotating_electronic_hamiltonian}) is
block diagonal with each block corresponding to a given
$\Omega_{u/g}^{(\pm)}$ subspace. As an example, let us consider the
subspace $0_u^+$. It is of dimension four. Figure \ref{fig:
Ilustration_AntiCrossings} illustrates the physical reason why a purely
long-range well arises in this subspace of states. If we consider only
the dipole-dipole interaction, one eigenvalue is purely repulsive,
while the three others are purely attractive, two of them being
identical (Figure \ref{fig: Ilustration_AntiCrossings}-a). They all
have the same asymptote. The four corresponding eigenstates are pure
Hund's case (a) states.  Let us consider separately the repulsive state
and the manifold of attractive states. If we ``turn on" the fine
structure coupling inside each of these two subspaces of states, while
neglecting the couplings between them, then the potential curves repel
each other and the asymptotes no longer coincide. Of course, since the
neglected couplings are not small, the four asymptotes have no
straightforward physical meaning. However, the important point is that
a crossing shows up between the repulsive curve and one attractive
curve (Figure \ref{fig: Ilustration_AntiCrossings}-b). Finally, if we
turn on the neglected fine structure terms, we couple the subspaces
corresponding to the two crossing states, and an anti-crossing appears
(Figure \ref{fig: Ilustration_AntiCrossings}-c).  The resulting
potential well is thus a consequence of the fine-structure mixing of
long-range molecular interactions, which links the inner, repulsive
dipole-dipole curve with an outer, attractive one.  What is remarkable
about this well is that {\it even the repulsive part occurs at very
long-range}, in a region where the asymptotic dipole-dipole expression
remains a very good approximation.  That is why the perturbative
approach used here is very well suited to describe the bound states
lying in this kind of well, or the so-called purely long-range
molecular states \cite{Stwalley}.

\begin{figure}[b]
\begin{center}
\includegraphics[height=6cm]{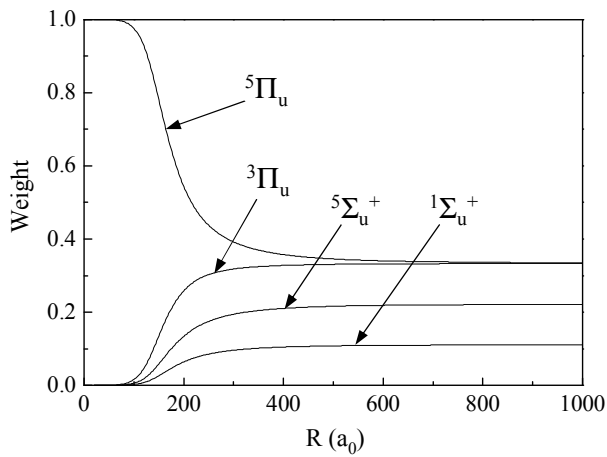}
\end{center}
\caption{\footnotesize Eigenstate for the $0_u^+$ purely long-range
potential well connected to the $2^3S+2^3P_0$ asymptote within the
fixed-nuclei approximation. The electronic state is given with its
decomposition in the Hund's case (a) basis set: the weights are the
squares of the projection on the different subspaces of Hund's case (a)
states. Distances are in atomic units.} \label{fig: Projections_Ouplus}
\end{figure}

Due to the competition between the dipole-dipole and the fine structure
interactions, not only the potential curves, but also the electronic
states explicitly depend on $R$. As an illustration, the $0_u^+$ purely
long-range electronic eigenstate is shown in Figure \ref{fig:
Projections_Ouplus}. The eigenstate is given with its projections over
the Hund's case (a) basis set. It evolves from the pure Hund's case (a)
$^5\Pi_u$ at short range, where the dipole-dipole interaction
dominates, to a pure Hund's case (c) for asymptotically large values of
$R$ where the dipole-dipole interaction vanishes like $1/R^3$.
Consequently, the fixed-nuclei approximation must be corrected by an
accounting of the coupling between the electronic and nuclear degrees
of freedom.
\\

The discussion just presented can also be applied to all the other
$\Omega_{u/g}^{(\pm)}$ subspaces. Figure \ref{fig:Non_Rotating} shows
three purely long-range {\it ungerade} potential wells. One is
connected to the $2^3S_1+2^3P_0$ asymptote and belongs to the $0_u^+$
subspace; it has been presented above. The two others are connected to
the $2^3S_1+2^3P_1$ asymptote and belong to the $0_u^-$ and $2_u$
subspaces. Within the fixed nuclei approximation the calculated $0_u^+$
well is $2.130$ GHz deep, the $2_u$ one is 0.321 GHz deep, and the
$0_u^-$ one is 0.054 GHz deep. We will examine these wells more closely
in the following discussion.

\subsection{Description of the motion of the nuclei}

So far the dynamics of the electrons has been treated independently
from the dynamics of the nuclei. In our perturbative model, the
coupling between the two comes from the kinetic energy operator for the
relative motion of the nuclei:
\begin{eqnarray}\label{eq: nuclear_kinetic_energy}
\hat{T}_n(R,\theta,\varphi)=-\frac{\hbar^2}{2\mu}\left(
\frac{1}{R}\frac{\partial^2}{\partial R^2}R -
\frac{\vec{\ell}^{\;2}}{\hbar^2R^2}\right)\cdot
\end{eqnarray}
In this expression $(R,\theta,\varphi)$ are the spherical coordinates
of the fictitious particle of reduced mass $\mu$ associated with the
pair of nuclei, and $\vec{\ell}$ is the orbital angular momentum associated
with its rotation.

\subsubsection{Effect of the rotation}
\label{parag: Effect of rotation}

\begin{figure}[t]
\begin{center}
\includegraphics[height=13cm]{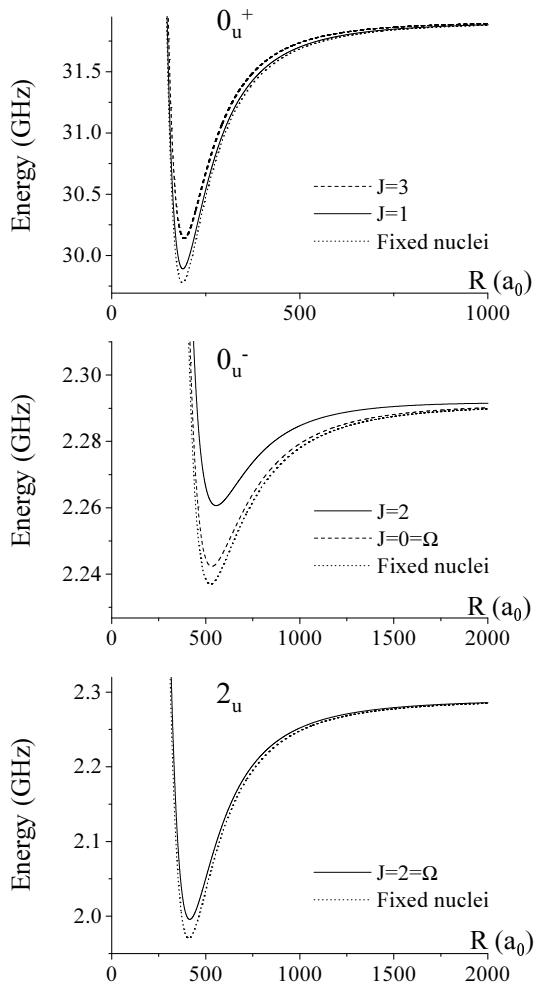}
\end{center}
\caption{\footnotesize Influence of the nuclear rotation on the
electronic potential energy for the three {\it ungerade} purely
long-range wells shown in Figure \ref{fig:Non_Rotating}. The dotted
lines are the result of the fixed nuclei approximation. The full lines
are the potential used to calculate the binding energies presented in
Table \ref{tab: results}. Note that the horizontal and vertical scales
are different for each graph.} \label{fig: Effect_Of_Rotation}
\end{figure}

First, the effect of the rotation of the nuclei on the electronic
states calculated above can be found if we add the last term of
(\ref{eq: nuclear_kinetic_energy}) to the hamiltonian (\ref{eq:
non_rotating_electronic_hamiltonian}). The operator to be diagonalized
becomes:
\begin{eqnarray}\label{eq: Rotating_electonic_states}
\hat{\mathcal{H}}=\hat{H}_{fs}(A) + \hat{H}_{fs}(B) +  \hat{U}(R) +
\frac{\vec{\ell}^{\;2}}{2\mu R^2}
\end{eqnarray}
Now, the space of states has to be extended to the rotational degrees
of freedom. Only the total angular momentum
$\vec{J}=\vec{L}+\vec{S}+\vec{\ell}$ has to be conserved
\footnote{Here, $\vec{L}$ and $\vec{S}$ represent the {\it molecular}
orbital and spin angular momenta.}, so we must consider the set of
states $\ket{\phi_{J,\Omega_{u}^{\pm}}}$ defined by the product of
electronic states determined above $\ket{\Omega_{u}^{(\pm)}}$ and of
rotational states $\ket{J,M,\Omega}$  \cite{Hougen} :
$\ket{\phi_{J,\Omega_{u}^{\pm}}} =\ket{\Omega_{u}^{(\pm)}} \otimes
\ket{J,M,\Omega}$. The quantum number $M$ is the projection of
$\vec{J}$ onto a lab-fixed frame. Since the molecule is linear,
$\vec{\ell}$ is orthogonal to the molecular axis, which means
$\ell_z=0$ and $J_z=L_z+S_z$. Thus the electronic quantum number
$\Omega$ represents the projection of $\vec{J}$ onto the molecular axis
and it is recalled as a parameter in the notation for the rotational
state. In this basis, $\vec{\ell}$ can be written as
$\vec{\ell}=\vec{J}-\vec{L}-\vec{S}$, the square of which is given by:
\begin{eqnarray}\label{eq:Detail l2}
\hat{\vec{\ell}}^{\;2}&=&\hat{\textbf{J}}^2+\hat{\textbf{L}}^2+\hat{\textbf{S}}^2-2\;\hat{J_z}^2+2\;\hat{L_z}\hat{S_z} +\;(\hat{L}_+\hat{S}_-+\hat{L}_-\hat{S}_+) \nonumber~\\
&\;&-\;(\hat{J}_+\hat{L}_-+\hat{J}_-\hat{L}_+)-(\hat{J}_+\hat{S}_-+\hat{J}_-\hat{S}_+)
\end{eqnarray}

In Equation (\ref{eq:Detail l2}), the first line contains terms that
couple electronic states with each other {\it inside} each
$\Omega_{u}^{(\pm)}$ block. The second line contains the terms that
couple states belonging to different $\Omega$ subspaces, due to the
action of $\hat{J}_{\pm}$ which obeys {\it anomalous} commutation rules
\cite{Zare} and couples $\Omega$ to $\Omega \mp 1$. These off-diagonal
coupling terms become important where potential curves belonging to
different $\Omega$ subspaces cross each other; they produce
anti-crossings. For the three purely long-range wells of interest, such
crossings appear far enough in the classically forbidden region that
the off-diagonal coupling terms can be neglected in the calculation of
the binding energy. Thus, in the following calculation, only the terms
coupling states within a given $\Omega$ subspace (first line in
Equation (\ref{eq:Detail l2})) are kept in the expression of the
rotation of the nuclei.

Figure \ref{fig: Effect_Of_Rotation} shows the change in the three {\it
ungerade} potential wells resulting from the inclusion of the rotation
of the nuclei in the hamiltonian. The minimum possible value for $J$ is
$J=\Omega$. For higher values of $J$ the contribution of the
centrifugal barrier due to the rotation of the nuclei increases.
Bose-Einstein statistics dictates that $J$ should be odd for $0_u^+$,
and even for $0_u^-$ (see e.g. \cite{Hougen}). There is no restriction
on $J$ for the $2_u$ state, since it is doubly degenerate.

\subsubsection{Effect of the vibration}
\label{section effect of vibration}

Next, since the electronic states depend on $R$ (Figure \ref{fig:
Projections_Ouplus}), the vibration of the nuclei also influences the
electronic degrees of freedom. This effect is described by the radial
part of the kinetic energy of the nuclei, namely the first term in
parenthesis in Equation (\ref{eq: nuclear_kinetic_energy}). This final
addition to the hamiltonian leads to the following equation:
\begin{eqnarray}\label{eq: bound states hamiltonian}
\hat{H}\ket{\psi} = \left\{-\frac{\hbar^2}{2\mu}\
\frac{1}{R}\frac{\partial^2}{\partial R^2}R
+\hat{\mathcal{H}}\right\}\ket{\psi}=E\;\ket{\psi}
\end{eqnarray}
where the eigenstates $\ket{\psi}$ are written using a basis with
separable variables: $\ket{\psi}=\ket{\chi_v} \otimes
\ket{\phi_{J,\Omega_{u}^{\pm}}}$, with $\ket{\chi_v}$ the vibrational
part, and $\ket{\phi_{J,\Omega_{u}^{\pm}}}$ the electronic and
rotational part. With these notations,
$\ket{\phi_{J,\Omega_{u}^{\pm}}}$ are the $R$-dependent eigenstates of
the hamiltonian $\hat{\mathcal{H}}$, with the eigenvalues
$V_{J,\Omega_{u}^{\pm}}(R)$ determined previously and given in Figure
\ref{fig: Effect_Of_Rotation}.

Because the crossings between electronic potential curves lie far
enough in the classically forbidden region, the action of
$\partial^2/\partial R^2$ on the electronic part should be considered
as a diagonal correction and we neglect the off-diagonal terms of this
operator. This is the so-called adiabatic approximation
\cite{Lefebvre-Brion}, and Equation (\ref{eq: bound states
hamiltonian}) reduces to a set of independent radial equations:
\begin{widetext}
\begin{eqnarray}\label{eq: radial equation}
\left\{-\frac{\hbar^2}{2\mu}\;\left(\frac{d^2}{dR^2}+\left\langle
\phi_{J,\Omega_{u}^{\pm}} \left| \frac{\partial^2}{\partial R^2}
\right| \phi_{J,\Omega_{u}^{\pm}} \right\rangle \right)+
V_{J,\Omega_{u}^{\pm}}(R) -E_{J,\Omega_{u}^{\pm},v}\right\}\;u(R)=0 \
\mbox{,}
\end{eqnarray}
\end{widetext}
where the vibrational part of the wave function has been written
$\bra{\vec{R}}\chi_v\rangle=u(R)/R$, and $E_{J,\Omega_{u}^{\pm},v}$ is
the binding energy for the ro-vibrational level $(J,v)$ in the
$\Omega_{u}^{\pm}$ potential well. Finally, the vibration of the nuclei
is described through a single effective potential well which is the sum
of $V_{J,\Omega_{u}^{\pm}}(R)$ (which already takes into account the
rotation of the nuclei) and of the correction coming form the
dependence in $R$ of the eigenstates of $\hat{\mathcal{H}}$.

\subsection{Calculation and comparison with the experimental spectrum}
\label{section: Theory Binding energies}

Table \ref{tab: comparison} provides the comparison between the
experimental results obtained for the $0_u^+$ potential well (column
A), and the calculated binding energies from the adiabatic approach
developed above (column B). In column (A), the measured binding
energies include the corrections discussed in Section \ref{section:
shifts}. Within the experimental accuracy, the agreement between our
measurement and our predictions for $J=1$ is remarkably good (except
for the $v=5$ line, which is too close to the atomic resonance to be
observed). Note that the $v=0$ line was probed with a different laser
set up, so its measured binding energy is less precise than the others
(see \cite{Leonard}). Also, the $(0_u^+,J=3)$ progression is too weak
to be observed in our experiment.

The effect of retardation on the calculated energy is illustrated by
the quantity $\epsilon^{Ret}$ (Table \ref{tab: comparison}, column C).
It increases the depth of the well, and therefore the binding energies
as well. Compared with the non retarded calculation ($k \rightarrow 0$
in the expressions \ref{eq:_Retarded_Dipole_1} and
\ref{eq:_Retarded_Dipole_2}), retardation is a correction proportional
to $R^2$ in relative value, but to $1/R$ in absolute value. Therefore,
it becomes very important in relative values for very elongated states
(up to $\sim 30 \%$ for $v=5$), and it is more important in absolute
values for less elongated states ($\epsilon^{Ret}=-6.6 MHz$ for $v=0$).
Given the experimental accuracy of 0.5 MHz, this work is a
demonstration of the retardation effect, which has to be taken into
account to reproduce the measured binding energies. This effect has
been already demonstrated for sodium atoms in 1996 \cite{Retard_Na}.

The correction to the electronic potential due to the vibration of the
nuclei is illustrated by the quantity $\epsilon^{Rad}$ in Table
\ref{tab: comparison}, column (D). Practically $\epsilon^{Rad}$ is the
difference between the binding energy calculated with and without the
term $\bra{\phi_{J,\Omega_{u}^{\pm}}}
\partial^2/\partial R^2 \ket{\phi_{J,\Omega_{u}^{\pm}}}$ in Equation
(\ref{eq: radial equation}). This term is part of the kinetic energy of
the system. Thus it brings a positive contribution to the effective
electronic potential and it moves the bound states upward in the wells.
Its contribution is non vanishing in the region where the electronic
state changes its character with $R$ due to the anti-crossings
discussed previously, that is to say in the vicinity of the bottom of
the potential well. Therefore the correction is stronger for the
deepest states, as they don't extend very far from this region. Weakly
bound states extend much farther into regions where the electronic
state does not depend strongly on $R$ (pure Hund's case c), and the net
effect is less pronounced.

\begin{table}[t]
\caption{\footnotesize Comparison between experimental and theoretical
binding energies in the case of the $0_u^+$ purely long-range potential
well. Column (A) gives the experimental results, after the corrections
discussed in Section \ref{section: shifts} are applied. Column (B)
gives the binding energy $E_{v,J}$ calculated from equation \ref{eq:
radial equation} within the adiabatic approximation. For each bound
state, $\epsilon^{Ret}$ is an estimate of the contribution to $E_{v,J}$
of the retardation effect. $\epsilon^{Ret}$ comes from the comparison
with the non-retarded calculation. Similarly, $\epsilon^{Rad}$ is the
calculated estimate of the term $\bra{\phi_{J,\Omega_{u}^{\pm}}}
\partial^2/\partial R^2 \ket{\phi_{J,\Omega_{u}^{\pm}}}$ (see Equation
(\ref{eq: radial equation})). Note that the binding energies presented
in column (B) already implicitly contain the contributions
$\epsilon^{Ret}$ and $ \epsilon^{Rad}$. All the energies are given in
units of $h$, in MHz.}
\begin{ruledtabular}
\begin{tabular}{cc|ccc}
  & (A) & (B) & (C) & (D) \\
  $v$ & Experiment & $E_{v,J}$ & $\epsilon^{Ret}$ & $ \epsilon^{Rad} $\\
\hline
 & & & & \\
 5 & - & -2.487 & -0.78 & +0.053 \\
4 & -18.2 $\pm 0.5$ & -18.12 & -1.6 & +0.28 \\
3 & -79.6 $\pm 0.5$ & -79.41 & -2.6 & +0.95 \\
 2 & -253.3 $\pm 0.5$ & -252.9 & -3.9 & +2.4 \\
1 & -648.5 $\pm 0.5$ & -648.3 & -5.2 & +5.3 \\
 0 & -1430 $\pm 20$ & -1418 & -6.6 & +10.3 \\
 & & & & \\

\end{tabular}
\end{ruledtabular}
\label{tab: comparison}
\end{table}

Finally, the high accuracy of the data and the good agreement between
the experimental and calculated spectra lead to an experimental
determination of the $C_3$ coefficient, which describes the
dipole-dipole interaction. In our calculations, changing $C_3$ by
$0.1\%$ changes the binding energies by at most 0.3 MHz, which is of
order of our experimental accuracy. Therefore, the present results
confirm the theoretical value used for the $C_3$ coefficient to within
$0.2\%$. As a consequence of Equation \ref{eq: C3}, we can infer that
the atomic radiative decay rate is $\Gamma=2\pi \times (1.625 \pm
0.003)$ MHz. As far as we know, this is the most accurate experimental
determination for the helium $2^3P$ decay rate.

\subsection{Other {\it ungerade} giant dimers}
\label{section: Other wells}

\begin{table}[b]
\caption{\footnotesize Results of the calculation detailed in the text
for the three purely long-range {\it ungerade} wells. Column (A) gives
the binding energy $E_{v,J}$ calculated within the adiabatic
approximation. The three last columns illustrate the unusual size of
the dimers. $R_{min}$ and $R_{max}$ are classical inner and outer
turning points, $\langle R \rangle$ is the mean internuclear distance.
All the energies are given in MHz, and the lengths in atomic units.}
\begin{ruledtabular}
\begin{tabular}{ccc|rrr}
 & & (A) & (B) & (C) & (D) \\
 & $v$ & $E_{v,J}$\footnote{Binding energies are given with respect to the asymptote of the potential considered.} & $R_{min}$ & $R_{max}$ & $\langle R \rangle$ \\
\hline
 & & & & & \\
$0_u^+$,  $J=1$& 5 & -2.487 & 147.6 & 2182 & 1797 \\
 & 4 & -18.12 & 147.7 & 1122 & 917 \\
 & 3 & -79.41 & 148.1 & 689 & 560 \\
 & 2 & -252.9 & 149.5 & 467 & 379 \\
 & 1 & -648.3 & 152.9 & 336 & 276 \\
 & 0 & -1418 & 162.5 & 246 & 213 \\
 & & & & &  \\
\hline
 & & & & &  \\
$0_u^-$,  $J=2$& 0 & -7.304 & 461.7 & 970 & 824 \\
 & & & & & \\
\hline
 & & & & & \\
$2_u$,  $J=2$& 3 & -4.584 & 320.5 & 2097 & 1712 \\
 & 2 & -21.41 & 322.5 & 1231 & 999 \\
 & 1 & -72.32 & 329.3 & 808 & 659 \\
 & 0 & -191.5 & 351.1 & 558 & 477 \\
 & & & & &  \\

\end{tabular}
\end{ruledtabular}
\label{tab: results}
\end{table}

Bound states in {\it ungerade} potential wells other than $0_u^+$ have
not been explored. However, the calculation presented above can also be
applied for those. Table \ref{tab: results} presents the theoretical
results for the molecular binding energies and characteristic sizes in
the three {\it ungerade} purely long-range potential wells. Column (A)
gives the results obtained when one solves Equation \ref{eq: radial
equation}. Experimentally, bound states are produced by driving an
electric dipole transition from the electronic state $^5\Sigma_g^+$
with $J=2$, so only $J=1$, 2 or 3 are accessible. In Table \ref{tab:
results}, the results are given for one relevant value of $J$, taking
into account the Bose-Einstein statistics already mentioned in
paragraph \ref{parag: Effect of rotation}.

The purely long-range character of these molecules arises from the very
large distance at which their inner classical turning points lie (Table
\ref{tab: results}, column B). The outer turning points (column C) and
mean sizes $\langle R \rangle =\bra{\chi_v}R\ket{\chi_v}$ (column D)
are also particularly large, leading to an unusual type of ``giant"
dimer for which asymptotic calculations allow an accurate description.
At such large distances, the next order term $C_6/R^6$ in the
electromagnetic interaction can clearly be neglected. The $C_6$
coefficient has never been published for this system, but one can
estimate that it is smaller than the value of $C_6=3265$ a.u. for the
$2^3S-2^3S$ interaction \cite{Starck} and calculate the order of
magnitude of the neglected term. For internuclear distances larger than
150 $a_0$, which is the range of interest for these purely long-range
molecules (see Table \ref{tab: results}), $C_6/R^6<C_3/R^3\times 1.5\
10^{-4}$. So neglecting this term leads to an error smaller than the
one due to the uncertainty on $C_3$.

While writing the present article we were informed that Venturi {\it et
al.} \cite{Venturi} had submitted for publication the result of a
multichannel calculation, which is also in very good agreement with our
experimental results. Their method is more elaborate and allows for a
direct solution of the full set of equations (\ref{eq: bound states
hamiltonian}). However, the binding energies obtained by both methods
are equal to within 0.5 MHz for all the bound states presented in Table
\ref{tab: results}. We have also performed a multi-channel resolution
of Equation \ref{eq: bound states hamiltonian} with the use of a mapped
Fourier grid method. Our results \cite{Proceeding_Australie} are
comparable to those of reference \cite{Venturi} to within 100 kHz. The
main reason why the adiabatic approach is efficient and the
multi-channel calculation required is that there is no crossing between
the adiabatic potential wells of interest and the other potential
curves. This allows for a single-channel calculation that leads to
Equation \ref{eq: radial equation} and is accurate enough to reproduce
the experimental spectrum.

\section{Summary and conclusion}

In a previous Letter \cite{Leonard}, we reported an accurate
measurement of the binding energies of purely long-range helium dimers
in the $0_u^+$ potential well connected to the $2^3S_1+2^3P_0$
asymptote. The present paper reports theoretical calculations which
complement the experimental results in order to interpret the spectra
measured.

The experiment consists in measuring the PA laser detunings for which a
strong heating of the atomic cloud is observed. The heating is assumed
to be a consequence of the resonant excitation of a bound state in the
$0_u^+$ potential well. To infer the corresponding binding energy, the
measured PA laser detunings must be corrected from a mean shift of the
molecular lines due to the non-zero magnetic field $B_0$ at the center
of the trap, and also to the non-zero temperature of the cold gas.
Since the detunings are measured with high accuracy, a simple
calculation shows that the temperature-induced shift must be
considered, given the range of temperature explored (2 - 30 $\mu$K).
This calculation does not include the exact shape and width of the
lines but only gives in a mean correction. The binding energies deduced
after correction are independant of the density, and no magnetic dipole
moment is detectable for the excited state. Apart from the symmetric
and asymmetric broadening mechanisms discussed in Section \ref{section:
shifts}, the lineshapes are actually also influenced by the dynamics of
the heating mechanism. Indeed the temperature curves are an indirect
measurement of the lineshape which relies on the efficiency of the
thermalization of the cloud. An incomplete thermalization can lead to
another source of broadening of the lines, but no additional shift. The
calorimetric detection scheme and its implications on the lineshape
will be discussed in a separate paper.

Here we have presented an approximate solution of the Schr\"odinger
equation that is well suited for asymptotically large internuclear
distances. The adiabatic approach allows for accurate calculations of
the binding energies in the case of purely long-range potential wells.
The calculation can easily be extended to other purely long-range
potential wells which can in principle also be observed in our
experimental conditions, namely $0_u^-$ and $2_u$.

Finally, the comparison between the experimental and theoretical
determination of the binding energies in the $0_u^+$ potential well is
very good if retardation effects are taken into account. As a
consequence, an accurate measurement of the radiative decay rate for
the excited atomic state $2^3P$ can be inferred. The accuracy of the
experimental data allows for a test of retardation effects as well as
of tiny vibration-induced couplings between electronic and nuclear
degrees of freedom.

Thus, the excellent agrement between our perturbative calculation and
our experiment suggests a good understanding of the purely long range
system. This work is a first step towards a better knowledge of pair
interactions in ultra-cold metastable helium. Further developments will
follow in order to measure the s-wave scattering length for two atoms
interacting through the $^5\Sigma_g^+$ electronic potential.

{\bf Acknowledgements :} The authors thank the group of F.
Masnou-Seeuws, at Laboratoire Aim\'e Cotton in Orsay, for fruitful
discussions.

\end{document}